
\input epsf                                                               %
\input harvmac
\noblackbox
\def\Title#1#2{\rightline{#1}\ifx\answ\bigans\nopagenumbers\pageno0\vskip1in
\else\pageno1\vskip.8in\fi \centerline{\titlefont #2}\vskip .5in}

%
%
\ifx\epsfbox\UnDeFiNeD\message{(NO epsf.tex, FIGURES WILL BE IGNORED)}
\def\figin#1{\vskip2in}
\else\message{(FIGURES WILL BE INCLUDED)}\def\figin#1{#1}
\fi
\def\Fig#1{Fig.~\the\figno\xdef#1{Fig.~\the\figno}\global\advance\figno
 by1}
%
%
%
%
\def\ifig#1#2#3#4{
\goodbreak\midinsert
\figin{\centerline{\epsfysize=#4truein\epsfbox{#3}}}
\narrower\narrower\noindent{\footnotefont
{\bf #1:}  #2\par}
\endinsert
}
%
%

\def\Hsl{{\,\raise.15ex\hbox{/}\mkern-12mu H}}
\font\ticp=cmcsc10
\def\RN{Reissner-Nordstrom}

\def\calh{{\cal H}}
\def\calv{{\cal V}}

\def\ajou#1&#2(#3){\ \sl#1\bf#2\rm(19#3)}
\def\frac#1#2{{#1 \over #2}}

\def\hf{{1\over2}}

\def\rst{r_*}
\def\eg{{\it e.g.}}

\def\p+{{\partial_+}}

\def\delsl{\,\raise.15ex\hbox{/}\mkern-13.5mu \nabla}
\def\Ssl{{\,\raise.15ex\hbox{/}\mkern-10.5mu S}}

\def\calo{{\cal O}}
\def\cala{{\cal A}}
\def\cald{{\cal D}}

%
%
\lref\QETDBH{S.B. Giddings and W.M. Nelson, ``Quantum emission from
two-dimensional black holes,''\ajou Phys. Rev. &D46 (92) 2486,
hep-th/9204072.}
\lref\Sredpc{M. Srednicki, private communication.}
\lref\ItZu{C. Itzykson and J.-B. Zuber, {\sl Quantum Field Theory} (McGraw
Hill, 1980).}
\lref\DGKT{F. Dowker, J.P. Gauntlett, D. Kastor, and J. Traschen, ``Pair
creation of dilaton black holes,'' Chicago preprint EFI-93-51,
hep-th/9309075.}
\lref\DGGH{F. Dowker, J.P. Gauntlett, S.B. Giddings, and G. Horowitz, to
appear.}
\lref\Erns{F. J. Ernst, ``Removal of the nodal singularity of the
C-metric,'' \ajou J. Math. Phys. &17 (76) 515.}
\lref\Suss{L. Susskind, ``String theory and the principles of black hole
complementarity,'' Stanford preprint SU-ITP-93-18, hep-th/9307168\semi
L. Susskind and L. Thorlacius, ``Gedanken experiments involving black
holes,'' Stanford preprint SU-ITP-93-19, hep-th/9308100\semi
L. Susskind, ``Strings, black holes and Lorentz contraction,''
Stanford preprint SU-ITP-93-21, hep-th/930813.}
\lref\Mart{E. Martinec, ``The light cone in string theory,'' Chicago
preprint EFI-93-21, hep-th/9304037.}
\lref\Lowe{D.A. Lowe, ``String causality,'' UCSB preprint UCSBTH-93-34,
hep-th/9310009.}
\lref\CBHR{S.B.~Giddings, ``Constraints on black hole remnants,'' UCSB preprint
UCSBTH-93-08, hep-th/9304027.}
\lref\EBHPP{D. Garfinkle, S.B. Giddings, and A. Strominger, ``Entropy in
Black Hole Pair Production,'' UCSB preprint UCSBTH-93-17,
gr-qc/9306023, to appear in {\sl
Phys. Rev. D}. }
\lref\tHoo{G. 't Hooft, ``Fundamental aspects of quantum theory related to
the problem of quantizing black holes,''  in {\sl Quantum Coherence}, ed.
J.S. Anandan (World Sci., 1990), and references therein\semi
Talk at 1993 ITP Conference, Quantum
Aspects of Black Holes, and private communication.}
\lref\BHQP{S.B. Giddings, ``Black holes and quantum predictability,''
UCSB preprint
UCSBTH-93-16, hep-th/9306041, to appear in the proceedings of the 7th
Nishinomiya-Yukawa Memorial Symposium, K. Kikkawa and M. Ninomiya, eds.}
\lref\Page{D. Page, ``Information in black hole radiation,'' Alberta
preprint  ALBERTA-THY-24-93, hep-th/9306083.}
\lref\HVer{K. Schoutens, H. Verlinde, and  E. Verlinde, ``Quantum black hole
evaporation,'' Princeton preprint PUPT-1395,
 hep-th/9304128\semi H. Verlinde, talk at 1993 ITP Conference, Quantum
Aspects of Black Holes.}
\lref\Liu{J. Liu, ``Evolution of pure states into mixed states,'' Stanford
preprint SU-ITP-93-1, hep-th/9301082.}
\lref\Gibb{G.W. Gibbons, ``Quantized flux tubes in Einstein-Maxwell theory
and  noncompact internal
spaces,'' in {\sl Fields and geometry}, proceedings of
22nd Karpacz Winter School of Theoretical Physics: Fields and
Geometry, Karpacz, Poland, Feb 17 - Mar 1, 1986, ed. A. Jadczyk (World
Scientific, 1986).}
\lref\AGGo{L. Alvarez-Gaum\'e and C. Gomez, ``Remarks on quantum
gravity,''\ajou Comm. Math. Phys.& 89 (83) 235.}
\lref\deAl{S.P. deAlwis, ``Quantization of a theory of 2d dilaton
gravity,''\ajou Phys. Lett. &B289 (92) 278, hep-th/9205069\semi
``Black hole physics from Liouville theory,''\ajou
Phys. Lett. &B300 (93) 330, hep-th/9206020.}
\lref\StTr{A. Strominger and S. Trivedi, ``Information consumption by
Reissner-Nordstrom black holes,'' ITP/Caltech preprint
NSF-ITP-93-15=CALT-68-1851, hep-th/9302080.}
\lref\GaSt{D. Garfinkle and A. Strominger, ``Semiclassical Wheeler wormhole
production,''\ajou Phys. Lett. &B256 (91) 146.}
\lref\VeVe{E. Verlinde and H. Verlinde, `` A unitary S-matrix for 2d black
hole formation and evaporation,'' PUPT-1380=IASSNS-HEP-93/8.}
\lref\Schw{J. Schwinger, ``On gauge invariance and vacuum
polarization,''\ajou Phys. Rev. &82 (51) 664.}
\lref\AfMa{I.K. Affleck and N.S. Manton, ``Monopole pair production in a
magnetic field,''\ajou Nucl. Phys. &B194 (82) 38.}
\lref\AAM{I.K. Affleck, O. Alvarez, and N.S. Manton,
``Pair production at strong
coupling in weak external fields,''\ajou Nucl. Phys. &B197 (82) 509.}
\lref\BOS{T. Banks, M. O'Loughlin, and A. Strominger, ``Black hole remnants
and the information puzzle,'' hep-th/9211030, {\sl Phys. Rev. D} to appear.}
\lref\Morg{D. Morgan, ``Black holes in cutoff gravity,"\ajou Phys.
Rev. &D43 (91) 3144.}
\lref\BaOl{T. Banks and M. O'Loughlin, ``Classical and quantum production
of cornucopions at energies below $10^{18}$ GeV,''\ajou
Phys.Rev. &D47 (93) 540.}
\lref\DXBH{S.B. Giddings and A. Strominger, ``Dynamics of Extremal Black
Holes,''\ajou Phys. Rev. &D46 (92) 627, hep-th/9202004.}
\lref\GHS{D. Garfinkle, G. Horowitz, and A. Strominger, ``Charged black holes
in string theory,''\ajou Phys. Rev. &D43 (91) 3140, erratum\ajou Phys. Rev.
& D45 (92) 3888.}
\lref\GiMa{G.W. Gibbons and K. Maeda, ``Black holes and membranes in
higher-dimensional theories with dilaton fields,''\ajou Nucl. Phys. &B298
(88) 741.}
\lref\Pres{J. Preskill, ``Do black holes destroy information?'' Caltech
preprint CALT-68-1819, hep-th/9209058.}
\lref\ACN{Y. Aharonov, A. Casher, and S. Nussinov, ``The unitarity
puzzle and Planck mass stable particles,"\ajou Phys. Lett. &B191 (87)
51.}
\lref\BHMR{S.B. Giddings, ``Black holes and massive remnants,''\ajou Phys.
Rev. &D46 (92) 1347, hep-th/9203059.}
\lref\Erice{S.B. Giddings, ``Toy models for black hole evaporation,''
UCSBTH-92-36, hep-th/9209113, to appear in the proceedings of the
International Workshop of Theoretical Physics, 6th Session, June 1992,
Erice, Italy.}
\lref\RSTii{J.G. Russo, L. Susskind, and L. Thorlacius, ``The
Endpoint of Hawking Evaporation,''\ajou Phys. Rev. &D46 (92) 3444.}
\lref\HawkTd{S.W. Hawking, ``Evaporation of two dimensional black holes,''
CalTech preprint CALT-68-1774, hep-th@xxx/9203052.}
\lref\CGHS{C.G. Callan, S.B. Giddings, J.A. Harvey, and A. Strominger,
``Evanescent black holes,"\ajou Phys. Rev. &D45 (92) R1005.}
\lref\RST{J.G. Russo, L. Susskind, and L. Thorlacius, ``Black hole
evaporation in 1+1 dimensions,''\ajou  Phys. Lett. &B292 (92) 13.}
\lref\BiCa{A. Bilal and C. Callan, ``Liouville models of black hole
evaporation,''\ajou Nucl.Phys. &B394 (93) 73, hep-th/9205089.}
\lref\BDDO{T. Banks, A. Dabholkar, M.R. Douglas, and M O'Loughlin, ``Are
horned particles the climax of Hawking evaporation?'' \ajou Phys. Rev.
&D45 (92) 3607.}
\lref\HawkEvap{S.W. Hawking, ``Particle creation by black
holes,"\ajou Comm. Math. Phys. &43 (75) 199.}
\lref\HawkUnc{S.W. Hawking, ``The unpredictability of quantum
gravity,''\ajou Comm. Math. Phys &87 (82) 395.}
\lref\BPS{T. Banks, M.E. Peskin, and L. Susskind, ``Difficulties for the
evolution of pure states into mixed states,''\ajou Nucl. Phys. &B244 (84)
125.}
\lref\Bank{T. Banks, ``TCP, quantum gravity, the cosmological constant and
all that,''\ajou Nucl.Phys. &B249 (85) 332.}
\lref\Hart{J.B. Hartle, ``Quantum kinematics of space-time. II: a model quantum
cosmology with real
clocks,''\ajou  Phys. Rev. &D38 (88) 2985.}
\lref\Unpub{S.B. Giddings, unpublished.}
\lref\HoWi{C.F.E. Holzhey and F. Wilczek, ``Black holes as elementary
particles,''\ajou Nucl.Phys. &B380 (92) 447.}

\Title{\vbox{\baselineskip12pt\hbox{UCSBTH-93-35}\hbox{hep-th/9310101}
}}
{\vbox{\centerline {Comments on Information Loss and Remnants}
}}
\centerline{{\ticp Steven B. Giddings}\footnote{$^\dagger$}
{Email addresses:
giddings@denali.physics.ucsb.edu, steve@voodoo.bitnet.}
}
\vskip.1in
\centerline{\sl Department of Physics}
\centerline{\sl University of California}
\centerline{\sl Santa Barbara, CA 93106-9530}

\bigskip
\centerline{\bf Abstract}

The information loss and remnant proposals for resolving the black hole
information paradox are reconsidered.  It is argued that in typical cases
information loss implies energy loss, and thus can be thought of in terms
of coupling to a spectrum of ``fictitious'' remnants.  This suggests
proposals for information loss that do not imply planckian energy
fluctuations in the low energy world.
However, if consistency of gravity prevents energy non-conservation, these
remnants must then be considered to be real.  In either case, the
catastrophe corresponding to infinite pair production remains a potential
problem.  Using \RN\ black holes as a paradigm for a theory of remnants, it
is argued that couplings in such a theory may give finite
production despite an infinite spectrum.  Evidence for this is found in
analyzing the instanton for Schwinger production; fluctuations
from the infinite number of states lead to a divergent stress
tensor, spoiling the instanton calculation.  Therefore na\"\i ve arguements
for infinite production fail.

\Date{}

\newsec{Introduction}

Although there are many variants\foot{For reviews see
\refs{\BHMR\Pres-\BHQP}.}, the three basic proposals for resolving the
problem of information loss in black holes are fundamental information
loss, remnants, or information return in the Hawking radiation.
Each of these possibilities has posed serious conceptual problems, and much
effort has been expended trying to overcome the difficulties for at least
one of these scenarios.  Two-dimensional models for black hole formation
and evaporation\refs{\CGHS} have recently served as a useful testing
ground for
these ideas.

In particular, two-dimensional black holes strongly suggest\refs{\QETDBH}
that information return
is unlikely without some new locality-violating physics. The basic
argument for this rests on treatment of the two-dimensional theories in a
$1/N$ expansion, where $N$, the number of matter fields, is large.
For the information to get out the rate of information
return from the black hole should be comparable its rate of energy loss for
the latter part of its evolution\refs{\Page}.  This includes a substantial
fraction of the lifetime of the black hole, where the $1/N$ approximation
would appear valid.  The energy flow is seen to leading order in the $1/N$
expansion, but the information flow is not, suggesting that it is supressed
by higher powers in $1/N$.  If this is the case information is not returned
until late in the evaporation, in the analog of the Planck regime,
when the expansion fails.

There have been several responses to this.  One suggestion is that the
$1/N$ expansion breaks down\refs{\Page,\HVer}.  One argument for this
is that
fluctuations in the vicinity of the horizon become strong, and this
invalidates the semiclassical reasoning\refs{\HVer}.  This contention
relies
in part on the assertion that if Hawking particles are traced back to the
vicinity of the horizon then they have near infinite frequency as seen by a
freely-falling observer.  However, it is not clear why it is valid to do
so.  In particular, if one examines the origin of the Hawking flux, for
example by explicit computation in the soluble
models of \refs{\BiCa\deAl-\RSTii}, then it is found that the Hawking
radiation
actually originates substantially outside the horizon where the trace
anomaly becomes important.  This corresponds to the known result in four
dimensions that the source of the Hawking radiation can not be localized
more precisely than the wavelength of the radiation, which is approximately
given by the radius of the hole.

Another response is to conjecture some new
type of fundamental non-locality in the
laws of physics.  One such conjecture is that of 't Hooft\refs{\tHoo},
who proposes that
information within a given volume can be determined by measurements on the
boundary of that volume.  He suggests that this could happen if the fundamental
laws of physics have some features similar to cellular automata.  This
would be interesting if a workable set of such laws were to be exhibited.
Alternately, Susskind has advocated the viewpoint\refs{\Suss}
that string theory has
precisely the right kind of non-locality built into it, basically from the
fact that if you try to measure a string on a very short time interval then
it spreads out.  He argues that when a string falls into a black hole,
observations of the external observer are effectively performing this type
of measurement and therefore cannot resolve the location of the information
on a scale less than the horizon size.  One objection to this is that it is
not clear what measurement
can actually be performed by an outside observer
to demonstrate that the string is indeed spreading out in the desired
way.\foot{One way to make
measurements on an infalling string that have reasonable resolution in
Schwarzschild time is to drop in, alongside the string, a particle
accelerator that is probing the string with higher and higher energy
particles as the string approaches the horizon.  These particles can then
be observed at infinity.  The collisions with these
probes spread the string out.  But in the absence of this arrangement
one is limited to observing whatever radiation is emitted from the
infalling string, and this will not give the desired time resolution.  In this
case it appears unnecessary to conclude that the string is spread out.}
Furthermore, there have been recent studies of causality in string
theory\refs{\Mart,\Lowe}.  These investigations suggest that string
causality is
not radically different from that in field theory, even at high energies.
It is likely that extension of these ideas could be used to show that
string theory does not allow the types of causality violation needed to get
the information out of the black hole.
This paper will take the point of view that such non-localities are not the
solution.

Instead the focus will be information loss and remnants.  Not long after
Hawking proposed information loss\refs{\HawkUnc} by generalizing the
S-matrix to an $\Ssl$ matrix acting on density matrices, it was
argued\refs{\BPS}  that
such evolution typically violates energy conservation, in so doing
violently disrupting low-energy physics.  In section three this paper revisits
this argument, and shows that in fact the $\Ssl$ matrices considered in
\BPS\ can be obtained through couplings to a hidden internal Hilbert space
of oscillators, at infinite temperature,
with which the Universe can exchange energy and information.   This in turn
suggests other proposals for information loss based on more general hidden
Hilbert spaces.  One particular possibility is a Hilbert space of
fictitious Planck-mass remnants.  This provides an example of an information
loss scenario that doesn't necessarily cook low energy physics.  This
scenario does, however, share with real remnants a problem of catastrophic
loss of energy through the analog of infinite pair production.  Whether one
views remnants as real or fictitious, this problem requires solution.
\RN\ black holes may be examples of objects that have
infinitely  many states\foot{This has been convincingly argued in the
semiclassical approximation in \refs{\StTr}.}
but aren't infinitely produced, and it is therefore
suggested that they serve as a viable paradigm for a theory of remnants.
The
remainder of the paper is devoted to investigating this possibility.  In
particular, in an effective theory describing such remnants couplings to
the electromagnetic field may be far from minimal.
These couplings may well
depend sensitively on the internal state of the remnant in a way that
invalidates the argument for infinite production.\foot{This is in contrast
to assumptions used in some formulas in \refs{\CBHR}.}
Such behavior seems to occur when
one examines euclidean instantons describing the analogue of Schwinger
production.

The present paper does not represent a detailed proposal to resolve the
problem of information loss, as the form of couplings of the
electromagnetic field to \RN\ black holes or similar remnants is
not yet fully understood.
However, I feel that great progress will be made toward
solving the black hole
information paradox if we can find where the logic that got us into it
might fail, and even better,
if there is any modification of Planck
scale physics that averts it.
This paper is a suggestion of where our ignorance might
have allowed
a resolution of the black hole information paradox to go unnoticed.
I hope to return to the details of couplings in future work.

\newsec{The effective approach}

In its basic formulation the question of information loss refers to issues
involving strong spacetime curvature and planckian physics.  However, the
fundamental paradox is phrased in terms of classical geometries and a definite
notion of time.  This has lead some to guess that perhaps the resolution to
the paradox lies in proper treatment of quantum geometry and time.

This contention, however, would appear to miss the mark.  Let us consider
formulating the problem in terms of a fully quantum-mechanical treatment
based on the Wheeler-deWitt equation, or whatever replaces it in the true
theory of quantum gravity.  To make contact with ordinary physics one needs
a notion of time, and this is a notoriously thorny issue.  However, in
the present problem one has the advantage that all questions can be asked
within the context of asymptotically flat space.  This means that we can
put a physical clock at infinity and use it to define what is meant by
time\refs{\Bank,\Hart}.  If $T$ is the dynamical clock variable,
then the full WdW wavefunction of clock plus gravitating system is of
the form $\Psi[T, f, g]$ where $f$ indicates matter fields and $g$ the
metric.  Let $\calh_c$ and $\calh_u$ be the contributions to the WdW
operator corresponding to the clock and the rest of the Universe,
respectively; the WdW equation is
\eqn\wdweq{\calh_{\rm WdW}\Psi=(\calh_c+\calh_u)\Psi=0\ .}
Consider arbitrary solutions $\psi$, $\phi$ of the equations
\eqn\wdws{\eqalign{ i {\partial \over\partial t} \phi &= \calh_c \phi\cr
i {\partial \over\partial t} \psi &= \calh_u \psi\ .}}
The dependence of $\calh_c$ on the variables $g,f$ can be taken to be very
weak by taking the clock to be very far away and very massive.  In this
approximation, \wdweq\ is separable and its general solution takes the form
\eqn\wdwsol{\Psi[T, f, g]= \int_{-\infty}^\infty dt \phi(T,t) \psi[t,f,g]\
.}
If the clock is a good one, then $\phi$ will be sharply peaked about $t=T$;
this can be arranged by taking the mass $M$ of the clock large.
The solution then becomes
\eqn\wdapp{ \Psi[T,f,g] \approx \psi[T,f,g] +\calo\left({1\over M}\right)}
and satisfies a Schr\"odinger equation,
\eqn\wdwapp{i{\partial\over\partial T} \Psi[T,f,g] = \calh_u
\Psi[T,f,g]+\calo\left({1\over M}\right) \ . }

Suppose that an asymptotic observer using a time slicing specified by this
clock watches diffuse dust collapse to form a black hole, and then
observes the decay products from the resulting evaporation.  The important
question is whether this observer sees the scattering to be unitary or not.
If it is unitary, one would like to know how and when the information came
out.  If it is not, one would like to have an effective description of what
generalization of the $S$-matrix maps the observer's initial state to the
final state.  In such a framework where
the black hole formation and evaporation is thought of as a
scattering process, and questions
formulated in terms of asymptotic observations, it is hard to see how the
answers could possibly get
mixed up in the subtleties of time or spacetime fluctuations.  Either the
bottom-line $S$-matrix is unitary or we would like to know what replaces
it.

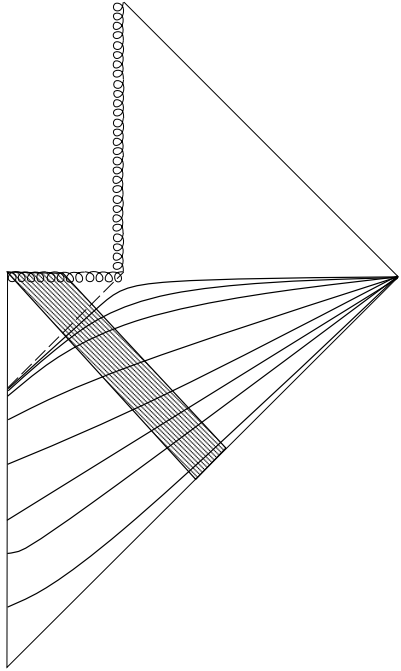
\ifig{\Fig\slicing}{Shown is a time slicing that avoids the interior of the
black hole.  It is plausible that evolution on this time slicing is
hamiltonian
until it reaches the planckian region near the classical
singularity.}{cilar.fig1}{3.5}

While on the topic of time, it can also be pointed out that the flexibility
in chosing time slicings in quantum gravity might be used as an advantage
in studying black hole formation and evaporation.  In particular, suppose
that one performs the exact quantization of the theory
choosing one's time slicing to always stay outside what is in the
semiclassical theory the black hole horizon, as indicated in fig.~1.  This
allows one to avoid the region of planckian curvature until the end of the
evaporation process.  Using such time slices suggests that at least up
until the endpoint of evaporation the process can be described in terms of
two coupled quantum systems.  The information thrown into the black hole
(and in correlation with the outgoing Hawking radiation) is all encoded in
the state on the left half of the timelike slice as it approaches the
horizon.  Of course the dynamics on these slices becomes more and more
extreme, and eventually involves planckian physics.  However, if one
believes that the only place that information is truly lost is the
singularity, then this can be avoided until the last instant of the
evaporation.

\newsec{Models for information loss}

Hawking proposed\refs{\HawkUnc} that information loss in quantum gravity
be described by a general linear evolution law for density matrices,
\eqn\doldef{\rho\rightarrow \Ssl\rho\ ,}
with $\Ssl$ a generalization of the usual $S$ matrix.  This proposal
was investigated in more detail by Banks, Peskin, and Susskind\refs{\BPS},
who within the context of $\Ssl$ matrices local in time
studied the constraints that the density matrix remain positive (so
as to have a probabilistic interpretation) and that the entropy be
non-decreasing.  If one considers a finite Hilbert space and
takes $Q^\alpha$ to be a complete set of hermitian matrices, the
infinitesimal form of \doldef\ is
\eqn\rdot{{\dot \rho} = \Hsl \rho=-i [H_0,\rho] -\hf\sum_{\alpha\beta\neq0}
h_{\alpha\beta} \left(Q^\beta Q^\alpha \rho + \rho Q^\beta Q^\alpha -
2 Q^\alpha \rho Q^\beta \right)\ ;}
here $H_0$ is the usual hamiltonian.  Ref.~\BPS\ argues that
sufficient conditions for positivity and increasing entropy are
that $h_{\alpha\beta}$ be positive and real, respectively.
Although it may be possible to construct other
physical $\Ssl$-matrices generated by an $\Hsl$,
these clearly represent a large fraction of the interesting
ones.\foot{For example, simple examples of $\Ssl$ matrices
preserving positivity but with
non-positive $h_{\alpha\beta}$ exist\refs{\Liu,\Unpub}.}

Eq.~\rdot\ can in fact be derived as the result of considering our
system to be in contact with another quantum-mechanical system which
is unobserved and therefore traced over.
Let the uncoupled hamiltonians of the two systems be $H_0$ and $H_h$,
with $[H_0,H_h]=0$.  Interactions between them arise from $H_i =
\sum_\alpha Q^\alpha \calo_\alpha$, where $\calo_\alpha$ are
operators acting on the ``hidden" Hilbert space,
and the total evolution is then
governed by
\eqn\hidcoup{H=H_0 + H_i + H_h\ .}

Consider first the case of a single $Q$, and let the internal system
be a harmonic oscillator of frequency $\omega$.  Take the coupling to be
of the form
\eqn\harmcoup{H_i= \sqrt{h\beta\omega\over 2} Q p_\omega}
where $p_\omega$ is the oscillator momentum.
Finally, let the harmonic oscillator be in a high-temperature
state,
\eqn\hdens{\rho_h=(1-e^{-\beta\omega})\sum_n e^{-\beta\omega n} |n\rangle
\langle n|\ ,}
with $\beta\rightarrow0$.
The density matrix of the observable system takes the form
\eqn\orho{\rho(t)={\rm Tr}_h\left( T e^{-i\int(H_0+ H_i) dt}\rho_h
\otimes\rho(0) T e^{i\int(H_0+
H_i)
dt}\right)\ ,}
where we work in the interaction picture for the internal system.
Expanding the exponential, we find for small times
\eqn\rhdota{\rho(\delta t) = \rho(0) - i \delta t [H_0,\rho(0)] -
h \int_0^{\delta t} dt \int_0^t dt' \beta\langle  [Q p(t),[Q p(t'),
\rho(0)]]\rangle_\beta\ .}
The thermal expectation value is easily computed,
\eqn\texpt{\eqalign{{\beta\omega\over 2}
\langle p(t)p(t')\rangle_\beta& = {\beta\omega
e^{-\beta\omega}\over
1-e^{-\beta\omega} } \cos\omega(t-t') + {\beta\omega\over 2}
e^{-i\omega(t-t')}\cr
&\buildrel{\beta\rightarrow0}\over\longrightarrow \cos\omega(t-t')\ ,}}
and we find
\eqn\rhevol{{\dot \rho} = -i[H_0,\rho] -h\int_0^t dt' cos(\omega
t') (Q^2\rho + \rho Q^2 - 2 Q\rho Q)\ .}
Therefore if we allow $Q$ to couple to an ensemble of oscillators
with a flat spectrum (that is we sum over all frequencies),
\rhevol\ becomes
\eqn\rhevola{{\dot \rho} = -i [H_0,\rho] - h (Q^2\rho + \rho Q^2
-2Q\rho Q)\ }
as in \rdot.

The generalization to multiple couplings is clear: simply diagonalize
$h_{\alpha\beta}$, and then introduce couplings to a family of
ensembles of oscillators labelled by $\alpha$.
The motivation for this construction is equally clear.  If one wishes
to reproduce \rdot\ through coupling to a hidden
quantum-mechanical system,
then that system should have a huge temperature so that it can raise
the entropy of the visible system indepent of its temperature.
However, to avoid the resulting infinite exchange of energy, the
coupling to the large-temperature system should fall with the inverse
temperature.  This limit furthermore has the desirable effect of
washing out correlations that arise between interactions of our
system with the hidden one at different times.  Finally, note that
positivity and reality of $h_{\alpha\beta}$ corresponds to
positivity of the norm in the hidden Hilbert space.

In the case of a field-theoretic model we wish to reproduce the
evolution law
\eqn\ftevol{\eqalign{{\dot \rho}= -i &\left[\int d^3x H_0(x),\rho\right]\cr
& - \int
d^3x d^3y
h_{\alpha\beta}(x-y)\left(\left\{Q^\beta(y)Q^\alpha(x),\rho\right\}
-2Q^\alpha(x) \rho Q^\beta(y)\right)\ .}}
This can likewise be done by couplings to a family of oscillator
ensembles.  For example, in the special case $h_{\alpha\beta}(x-y) =
h_{\alpha\beta} \delta^3(x-y)$, we simply need oscillators of all
possible frequencies at each point in space.  More generally one must
introduce correlations between oscillators at different points in
space, with correlation distance corresponding to the fall-off of
$h_{\alpha\beta}(x-y)$.

The above example suggests two points regarding information loss.
The first is that the evolution \rdot\ is readily extended to more general
descriptions of information loss that arise from couplings to more general
hidden systems.
The second is that if one expects
information to be lost during a definite time interval $\Delta t$,
this implies a corresponding loss of energy $\Delta E \sim 1/\Delta
t$.  A similar argument should hold (generalizing arguments of \BPS)
for information loss localized within a region of size $\Delta x$;
there corresponds a momentum loss $\Delta p\sim 1/\Delta x$.  To see
how these statements arise, consider for example restricting to frequencies
$\omega<\omega_0$.  Then \rhevol\ only reduces to \rhevola\ if we are
not capable of resolving times $\roughly<1/\omega_0$. At shorter
intervals non-trivial correlations appear, and
clustering fails.  Therefore if one restricts the energy loss
the information loss occurs over the corresponding time scale.\foot{A
sketch of a general
argument for this is as follows.  Consider a hamiltonian of the form
\hidcoup, and pass to interaction picture for the internal Hilbert space.
Then ${\dot \rho}= -i{\rm Tr}_h\left\{\left[H_0 + H_i(t),\rho\right]\right\}$.
Information loss is restricted to time interval $\Delta t$ if
${\rm Tr}_h\left\{\left[H_i(t),\rho\right]\right\}$ vanishes outside this
interval.  For this to happen in general, $H_i(t)$ should vanish outside
this interval.  This can only be arranged if the interactions connect
internal states with energies $\Delta E\roughly> 1/\Delta t$. }
Likewise, information loss
can only be localized to a region $\Delta x$ by making
$h_{\alpha\beta}(x-y)$ fall off at longer scales.  This implies that
it carries momenta $\calo(1/\Delta x)$.  If this is the largest
momentum loss, then information loss cannot be restricted to shorter
scales.

We can now consider more general types of unitarity violating evolution,
arising from coupling to various sorts of
quantum-mechanical systems.  Eq.~\rdot\ corresponds to unitarity violation
that is in a sense maximal.
In particular, we would like to know what is
likely to be the correct description of information loss for black
holes, if it is indeed lost.  First note that, following that argument at
the end of the preceding section, we might think that a correct
description is in terms of the Hilbert spaces describing states on the left
and right halves of the slices of fig.~1.  When the black hole finally
disappears, the Hilbert space on the left half of the slice becomes
inaccesible.  Therefore it is quite plausible that information loss in
black holes  be treated in terms of coupling to an internal Hilbert
space, as in \hidcoup,
which becomes invisible.  If this is the case we may not even care if there
is more fundamental non-unitarity at the singularity, as that dynamics
could be totally decoupled.
The hypothesis that black hole information loss can be described in terms
of coupling to an internal Hilbert space
fits nicely with the
reformulation of  Hawking's $\Ssl$ matrices in terms of such couplings, as
has just been outlined.  Alternatively black holes might
be described by more general forms of information loss arising from
different internal Hilbert spaces.
For example,
one might consider instead modelling their loss
by assuming the existence of a family of quantum fields\foot{A related
discussion appears in \refs{\AGGo}.} that couple
to ordinary quantum fields through operators that only become
important during the final stages of black hole evaporation.  These
could carry the black hole's information away.  Of course, in
principle this information might be recoverable in couplings through
the same operators that transferred it to the hidden Hilbert space.
However, this by no means implies that it is recoverable in practice,
as couplings to the hidden space may be small everywhere except in
black holes.
Indeed the reader may note that what is being discussed here is
nothing more than a theory of black hole remnants, in which the
remnants are not observed after they are produced.
Assuming that the information is truly lost in such a picture
corresponds to assuming that the remnants are fictitious -- nothing
more than bookkeeping devices to summarize the couplings through
which we lose information.  On the other hand, if the remnants are
real, then the information may just be hard to find.

Let us next reassess the logic of the information loss scenario.  As
shown in \BPS, information loss via an infinitesimally generated
$\Ssl$ matrix also violates
energy and momentum conservation.  As has been explicitly described,
such evolution corresponds to placing the world in contact with a
fictitious Hilbert space raised to infinite temperature.  This does not
agree well with observation.  However, one may consider more general,
and more innocuous,
forms of information loss.  The above arguments indicate a connection
between
information loss and energy loss.  An alternate model for
information loss is to imagine that
information is carried off by remnants that are fictitious in the
same sense as the many-oscillator Hilbert space.
These remnants also
carry away energy.  However, once we
have such a model we can eliminate the distastefulness of energy
non-conservation by instead assuming the remnants to be real!

It should be noted that in order to describe formation and
evaporation of near-Planck scale black holes we should consider
remnants with energies up to near the Planck scale.  A very plausible
assumption is therefore that black hole remnants have Planck-size masses.

Such remnant models (real or fictitious) of information loss
(temporary or permanent) clearly have a distinct advantage over
information loss via $\Ssl$-matrices:  they do not offer the
appearance of infinite temperature.  Suppose that the remnant state
is initially the vacuum.  With the assumption that the remnants have
Planck masses, any low energy scattering process that we perform
therefore does not couple to remnants through real processes,
i.e. does not see
information loss.  Virtual effects of remnants, although possibly
important,\foot{Related effects will be discussed in subsequent sections.}
do not lead to loss of information since every remnant line must
terminate in a closed loop.  On-shell remnants only enter once
scattering energies cross the  planckian threshold or once black holes are
formed.  Only in such cases is information lost.  There remains the
possibility that the information could be regained through subsequent
processes.  However, this could be made vanishingly unlikely in ordinary
circumstances if the operators to which the remnants couple only become
important at the Planck scale, and because if black holes are rare,
it is unlikely that a remnant from one black hole will reappear in
another.

Such models are not yet immune from problems.  Whether these
are considered models for information loss or true remnant scenarios,
one must have an infinite number of remnant species to carry off the
information from a black hole as large as you can imagine.  This
raises the standard objection to  remnants:  infinite
species seems to imply infinite total rates of production in any process
where there is enough available energy, e.g. inside the sun,
even if individual production rates are near infinitesimal.  In the
case where the remnants are considered fictitious this would
be interpreted as a catastrophic instability in which energy
disappears at an infinite rate.  These issues will be the focus of
much of the rest of the paper.

It should also be noted that since they carry energy, real remnants
could in principle be detected through their gravitational field.
Turning the logic around, this is yet another reason to believe that
remnants are real, rather than fictitious:  gravity seems
inconsistent in the absence of energy conservation.

In any case,
since aside from energy conservation
these models of information loss have the same features and
drawbacks whether or not the remnants are fictitious, it makes sense
drop the extra assumption of energy non-conservation and promote the
remnants to reality.  This will be done in the remainder of the
paper, although readers who prefer energy non-conservation may just as
well imagine the remnants fictitious.

\newsec{A remnant paradigm}

The preceding section has outlined a close relationship between remnants
and information loss.  $\Ssl$ matrices can be thought of as arising
from coupling to infinite fictitious
remnant species at infinite temperature.
More tame alternatives arise from a different infinite spectrum
of remnants, in its vacuum.  Although information loss is then in a
sense more palatable (and in a sense indistinguishable from
remnants), it still suffers the serious flaw corresponding to
infinite remnant production.

Information loss or remnants can be saved, and the black hole
information paradox solved, if an escape from this problem can be
found.  Although an ironclad escape has not yet been found, rather
strong suggestions arise by considering extremal \RN\ black holes.

If information is not returned in Hawking evaporation, then there
must be an infinite number of states of a \RN\ black hole of charge
$Q$.  These are formed by starting with any given extremal black hole
state, dumping in matter carrying arbitrary information, and then
allowing the black hole to evaporate back to extremality.
Real \RN\ black holes may well exist.  Furthermore, we do not observe
them to be infinitely pair produced.  This indicates that they
provide an excellent arena to investigate the
information paradox.\foot{Previous advocates of this include
\refs{\StTr}.}

Indeed, if information is not returned in Hawking radiation, then
\RN\ black holes appear to give an existence proof for objects with
all the desirable properties of remnants.  \RN\ black holes will
therefore be taken as a paradigm for a viable theory of information
loss/remnants.  We will seek to understand their essential properties
that allow them to fit this role.

\newsec{Effective theories for remnants}

In order to discuss issues of pair production and other effects of remnants
it is useful to have a general framework in which to describe them.  This
section will take steps towards constructing an effective theory for
remnants, and in particular will attempt to infer its general properties,
if such a theory exists.

Remnants and their interactions should be localized in spacetime.  Furthermore,
a theory of remnants should also be Lorentz invariant at long distances.
The only known (and possibly the only existing) way of reconciling
locality, causality, and Lorentz invariance in a quantum framework is
quantum field theory.  Therefore at distances large as compared to the remnants
and any of their interaction time scales, they should be described by a
field,
\eqn\rfield{I_A(x) = \int {d^3 k\over (2\pi)^3 2\omega_k}\left[ I_A(k)
e^{ik\cdot x} + I_A^\dagger(k)e^{-ik\cdot x}\right]\ ,}
where $A$ labels the different remnant states.
For simplicity we have assumed that the remnants do not carry spin,
 although this could be generalized.  The action governing free
propagation of a remnant should then be of the form
\eqn\informact{S_K = \int d^4x \sum_A \left[-\hf \left(\partial
I_A\right)^2 -\hf m_A^2 I_A^2\right]\ .}

Remnants also have couplings to the electromagnetic, gravitational, and
other fields.  To investigate their form, return to the case of \RN\ black
holes of charge\foot{To eliminate concerns of discharge by pair production one
may wish to consider magnetic charge, although the remaining discussion
does not depend on this.}
 $Q\gg1$, which we will represent by complex fields $I_A$.
First consider on which scales the dynamics can be described by
effective interactions.

To begin with, recall that the infinite degeneracy originates in the fact
that the extremal black hole could have been built out of matter
with any initial mass $M>Q$, which is then allowed to evaporate.
This gives an infinite number of possible initial states, and if
information is not returned in Hawking radiation the resulting extremal
hole has infinite states as well.  The evaporation time is of order $M^3$,
or very near extremality\StTr\ $Q^3$.  The resulting states may be
truly degenerate or only nearly degenerate.
Even if once the black hole nears
extremality it by some mechanism begins to radiate information, the time
required for all of it to escape is of order $M^4$, and the decay time
between the states is of order $M^2$.  Therefore for $M\gg Q$,
on time scales $\gg Q^3$ and $\ll M^2$ we have essentially stationary
configurations.

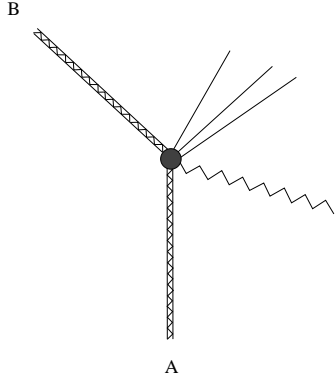
\ifig{\Fig\Remvert}{Shown is a typical vertex for
a photon interacting with a black
hole.  The photon is absorbed, but excites the black hole.  The black hole
then
de-excites by emitting some quanta, for example through Hawking radiation,
leaving it in a different internal state.}
{cilar.fig2}{2.0}

Now let us consider scattering electromagnetic radiation of frequency
$\omega\ll1/Q^3$ from the black hole.  On time scales $\gg Q^3$ the
process of absorption and reemission looks effectively pointlike, as
indicated in fig.~2.  Therefore we would expect that it be summarized by an
effective vertex operator at these scales.  This vertex describes both the
absorption of the incoming quantum and the reemission, by Hawking or other
process, of the energy which leaves the black hole back at extremality.  To
simplify the notation we
will assume the existence of a massless scalar field $f$ and will consider
only process in which the black hole absorbs a photon and emits quanta of
the scalar field.  (This saves writing lots of indices but makes no
essential change to the physics.)  The effective vertex for such a process
with $n$ quanta emitted is of the form
\eqn\effvert{\cala^\mu_{(n)BA}(p,k,p_i)\, A_\mu(k) \prod_{i=1}^n f(p_i)
I_B^\dagger(p)
I_A(p')\ .}
In general this will include a minimal coupling to the electromagnetic
field, although it is possible that $\cala^\mu_{(0)BA}(p,k)=0$, that is
the elastic scattering amplitude is zero.

These couplings will in general pair produce remnants, \eg\ via the
analogue\refs{\Gibb\GaSt-\EBHPP}
of the Schwinger process for production of \RN\ black holes in a
constant field.  The problem of infinite
production can be phrased as follows.  Suppose that we consider two
extremal black holes.  Suppose that they both were constructed by starting
with identical extremal holes, but that we have dropped the continent of
Africa into one of them and then waited a time $\gg\gg Q^3$ for its energy
to be reradiated and the black holes to settle down to states apparently
identical from the outside.  At first sight there is no obvious reason why
there wouldn't be equal production rates for these two types of black
holes, and by extension for infinitely many species.  We might describe
this by dividing the label $A$ into two sets, $a,\alpha$, where $\alpha$
parameterizes the infinite number of different states (e.g. corresponding to
things that were done to the black hole in the far past) that do not give
different vertices.  This means that \effvert\ becomes
\eqn\effverta{\cala^\mu_{(n)ba}(p,k,p_i)\, A_\mu(k) \prod_{i=1}^n f(p_i)
 I_{b,\beta}^\dagger(p)
I_{a,\alpha}(p')\ .}
The infinite degeneracy in the states labelled by $\alpha$ gives the
infinite production rate.
In particular, note that if the coupling is dominated by the minimal term
(i.e. the effective theory is weakly coupled as in \refs{\CBHR}), then it
is insensitive to the state and the production rate is infinite.

How is this problem to be avoided?  Due to difficulty in
deriving the effective description of \RN\ black holes, a concrete proposal
for the form of the couplings cannot yet be made.  However, one can make
some reasonable guesses as to what behavior is required and as to whether
it emerges.

In particular, note that implicit in the argument that \effverta\ is
independent of $\alpha$ was the assumption that we are working
on-shell or very nearby, and with real momentum.
Only in this context do the statements
about irrelevance of modifications of the black hole in the far past
apply.  However, in calculating pair-production rates for process far
below the Planck scale, one needs the couplings \effvert\ off shell or at
complex momenta.
It is quite conceivable that in these regions $\alpha$ independence
fails in a way that renders production finite.\foot{This is distinct from
the suggestion\refs{\BDDO,\BaOl} that the electromagnetic form factors
vanish at large momentum transfer, since for example Schwinger production
depends only on the form factors at small momentum transfers.}

One motivation for this is the observation that remnants should
involve Planck scale physics to describe them and their couplings to
other fields.  To see this consider forming one of our \RN\ remnants
by throwing a large mass into a black hole and allowing it to
evaporate.  The remnant state is what is left; in other words the
remnant can be described by taking the black hole plus Hawking
radiation and acting on it by a collection of annihilation operators
that eliminate the Hawking radiation.  If the resulting state is
evolved back in time, due to the absence of the Hawking radiation it
gets very singular in the vicinity of what was the horizon.\foot{This
has been emphasized by H. Verlinde \refs{\HVer} in a different
context.}  The resulting strong coupling and large modification of
the solution in the vicinity of the former horizon indicate that the
true remnant eigenstates have large support on configurations where
planckian physics is important.  This could well lead to the desired
strong dependence of the remnant couplings on the momenta.

Note also that vanishing of the elastic scattering amplitudes, and
thus of the
minimal coupling to the electromagnetic field, seems to
be required.  This prevents a non-vanishing amplitude for
Schwinger production with the internal state of the remnant
unexcited; this would be accompanied by an overall infinite factor.
It is quite plausible that the elastic amplitudes do indeed vanish.  To see
this, note that if we try to throw a photon of any energy at a black
hole, it can be absorbed and in doing so  excites the
internal state above extremality. This is followed by Hawking
radiation, for example of $f$ particles.  The dominance of these
processes (as opposed to off-shell elastic scattering) at momenta
where on-shell elastic scattering is not possible suggests that the
elastic amplitudes could in fact be zero.

To illustrate these comments, consider the problem of
the analogue of Schwinger pair production in this framework.
The decay rate can be computed\foot{See, \eg, \refs{\ItZu}.} from the
functional integral
\eqn\fctint{S_0[A_0^\mu] = \int \cald {\tilde A}_\mu \cald f \cald I
e^{iS[A_0^\mu +{\tilde A}^\mu, f, I_A]}\Biggr/
\int \cald {\tilde A}_\mu \cald f \cald I
e^{iS[{\tilde A}^\mu, f, I_A]} \ }
where the gauge field has been divided into background and fluctuation
pieces.  If $V_4$ is the four-volume in question, the rate is
\eqn\decay{V_4 \Gamma = -2 {\rm Re} \ln S_0[A_0^\mu]\ .}
In these expressions the action includes, in addition to the kinetic
piece \informact, coupling terms corresponding to the amplitudes
\effvert.  These take the position-space form
\eqn\intact{\eqalign {&\sum_{AB}\sum_n \int d^4x d^4x' d^4y \prod_i^n
 d^4z_i f(z_i)
I^*_B(x)
\cala^\mu_{n;BA}[x,x',y,z_i] I_A(x) A_\mu(y) \cr
&\equiv \sum_{AB}\int d^4x d^4x' I^*_B(x) {\calv}_{BA}[x,x';
A^\mu,f] I_A(x')\ . }}
In this equation $\cala^\mu_{n;BA}$ also may contain derivatives acting
on the fields, and in the first line we have supressed couplings to
multiple photon emission for simplicity.

The contribution of the (normalized) functional integral over $I$ to
\fctint\ is
\eqn\Ifnct{S_0[A,f] = {\rm Det}^{-1/2}\left\{\left[(-p^2-m_A^2
+i\epsilon)\delta_{AB}
+ \calv_{AB}\right]\biggr/(-p^2-m_A^2 +i\epsilon) \right\}\ ,}
with corresponding effective action $w$,
\eqn\effact{ Im\, \int d^4x w[A,f,x] = -2{\rm Re} \ln S_0[A,f]\ .}
The latter can be rewritten
\eqn\effacta{\eqalign{ Im\, \int d^4x w[A,f,x] = \int d^4x
Re\int_0^\infty {ds\over
s}e^{-s\epsilon} \sum_A& \langle x,A| e^{-is\left[(p^2 + m_A^2){\bf 1} -
{\bf \calv} \right]}|x,A \rangle\cr & -\langle x,A|
e^{-is(p^2 + m_A^2)}|x,A
\rangle\ ,}}
or, in momentum space,
\eqn\effacta{\eqalign{Im\, \int d^4x w[A,f,x] = &\int d^4x
Re\int_0^\infty {ds\over
s}e^{-s\epsilon} \int {d^4p\over (2\pi)^4}
{d^4 p'\over (2\pi)^4} e^{i(p-p')x}
\sum_A\cr& \langle p',A| e^{-is\left[(p^2 + m_A^2){\bf 1} -
{\bf\calv} \right]}|p,A \rangle -\langle p',A| e^{-is(p^2 + m_A^2)}|p,A
\rangle\ .}}
If the vertex $\calv$ corresponded merely to minimal coupling, then
\effacta\ can be evaluated by continuation into the complex plane.  The
answer arises from a sum of terms at euclidean momenta that correspond to
the Schwinger instantons, which are the euclidean orbits in the background
field.  This result is then acompanied by an infinite factor from the sum
over remnant states.  In the example of GUT monopole
production\refs{\AfMa,\AAM}, $\calv$
picks up corrections due to the structure of the monopole. However, these
are supressed by powers of $1/M$, where $M\sim 1/R_{\rm monopole}$ is the
scale for monopole excitations.  In the limit of weak background fields,
the contributions of these to the low-lying instantons will be supressed by
powers of $eE/M$.  However, with the \RN\ paradigm for remnants there is no
mass gap.  One can no longer make the argument that the
contributions of the non-minimal couplings are small, and as suggested
above they may in fact be dominant.   Continuation into the complex plane
is no longer guaranteed to produce the Schwinger saddlepoint.
Although an explicit example of such couplings is lacking, it
is quite possible that they strongly depend on the state label $A$.  With
such a dependence it is possible to supress infinite production.

Clearly it would be desirable to derive the effective
couplings, both on- and
off-shell and at complex momenta,
between external fields and \RN\ black holes.
This is a very difficult task.
An important check to make is that interaction with a black hole must
necessarily excite it; otherwise the minimal coupling is non-vanishing and
infinite production likely results\refs{\CBHR,\Sredpc}.  It may also
be true that
there is no standard effective field theory that describes such couplings.
In any case, in the absence of knowing a detailed effective theory,
one seeks other means to attack the pair production
problem.  Another approach is the study of gravitational instantons
describing the production process.

\newsec{Pair production via instantons}

An analogue of the Schwinger process for the pair production of \RN\
black holes in a background field is described\refs{\Gibb,\GaSt} by
the euclidean version of the Ernst metric\Erns.
The black hole produced by this instanton is near-extremal; in fact
it is just far enough above extremality so that its Hawking
temperature matches its acceleration temperature.  The states produced
are thus in equilibrium with the Unruh radiation.
The action for this
metric is finite and has been computed\refs{\EBHPP}; as expected it is
of the form $S_E = -\pi Q/B+\calo(Q^2)$.
The first term gives Schwinger's rate, and the second term contains
a contribution
that is precisely the black hole entropy
and suggests that the number of states being produced is $\exp\{S\}$.

However, it is clear that this is not the complete story.  In
particular, subleading corrections to the production rate also come
from the fluctuation determinant, and this might be expected to
incorporate the infinite number of states of the black hole.

Computing the full fluctuation determinant for arbitrary
gravitational, electromagnetic, and other excitations about the
instanton is a difficult problem.  However, two simplifications can
be made while retaining the essential flavor of the calculation.
First, we will consider fluctuations only in the spectator field $f$.
Secondly, we are clearly interested in fluctuations only near the
horizon.  In the limit of small external field, the instanton becomes
effectively two dimensional in a large neighborhood of the horizon
and at energies $\roughly<1/Q$.  We can thus see the essential issues
by considering the low-energy states, that is, reducing to the s-wave
sector so the problem is purely two dimensional.

To be more explicit, in the small $B$ limit the euclidean
Ernst metric in the
vicinity of the horizon takes the form
\eqn\Erlim{ds^2= Q^2(\sinh^2 y dt^2+ dy^2 + d\theta^2 + \sin^2 \theta
d\phi^2) \
.}
The $y,t$ part is a solution to the reduced action
\eqn\redact{ {1\over2\pi}\int d^2 \sigma \sqrt{g}\left\{
e^{-2\phi}\left[R+2(\nabla\phi)^2\right] + 2-2Q^2e^{2\phi}\right\}\ ,}
where $e^{-\phi}$ is the two-sphere radius.  The fluctuations of s-wave
part of the $f$ field will be weighted using
\eqn\fact{S_f= \hf \int  d^2 \sigma \sqrt{g} (\nabla f)^2\ }
where $f$ has been rescaled by a factor proportional to $1/Q$.

For big black holes we expect to be able to
work in the semiclassical limit and consider
such fluctuations on the fixed background.  Consider first quantizing them
in the canonical framework.  This is most easily done by introducing the
tortoise coordinate, $\rst$, in terms of which the two-metric is
conformally flat,
\eqn\rstmet{ds^2 = {Q^2\over \sinh^2 \rst} \left(dt^2 + d\rst^2\right)\ .}
Then the action \fact\ takes the flat-space form,
\eqn\ffact{S_f= \hf \int  d^2 \sigma_* \left[(\partial_t f)^2 + (\partial_*
f^2)\right]\ .}
The fluctuations are the usual left and right moving flat space modes, and
can be quantized by introducing the standard flat space inner product and
conjugate momentum.  The infinite number of states arises from the infinite
volume of $\rst$.  Transition amplitudes can alternatively be converted
into functional integrals by the standard procedure.  Throughout only the
flat metric $ds_*^2=g_{*ab} d\sigma^ad\sigma^b= dt^2 + d\rst^2$ enters, and
therefore the functional integral takes the form
\eqn\functint{\int\cald_{g_*}f e^{-S_f}\ .}
As has been explicitly indicated, since the quantization depends only on
$g_*$ one obtains the measure regulated with respect to $g_*$.
Eq.~\functint\ is infinite due to the infinite volume in $g_*$.
This could be interpreted as yielding the infinite factor in the  pair
production rate.

Note, however, that there is another quantization of the fluctuations that
gives a {\it finite} answer.  This arises if one starts with the functional
integral, but now regulated with respect to the euclidean continuation of
the physical metric,
\eqn\functreg{\int\cald_{g}f e^{-S_f}\ .}
The volume near the horizon as measured in the metric $g$ is finite, and
the divergent factor has been eliminated.

Which answer is correct, \functint\ or \functreg?  Note that the difference
between them is simply a conformal rescaling of the metric,
$g_*=\exp\{2\rho\}g$.  Conformal invariance of the action means that this only
affects the regulator.  Since we are
working in the two-dimensional limit, we can explicitly exhibit the
difference between the functional integrals in terms of the Liouville
action,
\eqn\lioud{\int\cald_{g_*}f e^{-S_f}= e^{S_L}\int\cald_{g}fe^{-S_f}\ ,}
with
\eqn\lioudef{S_L ={1\over24\pi} \int d^2\sigma \sqrt{g} \left[ (\nabla \rho)^2
+ R
\rho\right]\ .}
The difference in stress tensors can likewise be computed,
\eqn\strd{\eqalign{T_{z{\bar z}} =& T_{*z{\bar z}} + {1\over12}\del_z
\del_{\bar z} \rho\cr
T_{zz} =& T_{*zz} -{1\over12}\left( \del_z^2 \rho + (\del_z\rho)^2\right)\ .}}
The stress tensor $T$ corresponds to Hawking
radiation in the Hartle-Hawking state.  The difference between this and $T_*$
gives a divergent proper flux at the horizon, as in the difference with the
Boulware vacuum.  Similar behavior is expected
to occur more generally whenever there is a horizon.

The former corresponds to cutting off the fluctuations using
a cutoff in Kruskal momentum.  If on the other hand fluctuations
in the remaining infinite number of states are allowed, they make an
infinite contribution to the stress tensor near the horizon.
In this case the backreaction on the metric
becomes large and the semiclassical approximation breaks down.  This means
that in fact we had no right using the instanton to compute the production
rate for an arbitrary state among the infinite number of possible states in
the first place.  It isn't possible to tell if the total
production rate is finite or not -- to do so requires a more in-depth
calculation.  Because of the apparent divergence in the stress tensor this
could well involve Planck physics.

We therefore can't yet draw a concrete conclusion about the production
rate.  We can however see that the instanton calculation breaks down in a
way that suggests relevance of strong coupling physics in the vicinity of what
was the horizon.  This dovetails nicely with the observations made in the
preceding section; it is quite possible that this corresponds to couplings
to external fields that are very different from their on-shell, real
momentum values.  (It alternatley might indicate a breakdown of the effective
approach.)
This suggests that such a mechanism may prevent infinite production of \RN\
black holes.  And if such a mechanism works for \RN\ black holes, one may
conjecture that there exist other remnant models with the same properties.

\newsec{Comparison to the dilatonic case}

Other proposals for remnants with finite production have been made; notable
is the suggestion that extremal dilatonic black holes provide a model for
remnants\refs{\CGHS,\BDDO,\DXBH}, and that they have finite production
rates\refs{\BDDO,\BaOl,\BOS}.  The explanation proposed in
\BOS\ for finite production
is distinct from that proposed here.  In
particular, \BOS\ reasons that the rate is finite because the approximate
euclidean instanton describing pair production has finite volume, and thus
corresponds to production of a finite number of states. However, it has
subsequently been found that there are instantons corresponding to
production of infinite volume dilatonic black holes\refs{\DGKT,\DGGH}.

Furthermore, note that merely finite volume is not necessarily sufficient
to guarantee finite production.  This can be illustrated with the case of
\RN\ black holes, which also have an infinite spatial volume at
extremality.
This is not, however, the origin of the infinite number of presumed
states; slightly above extremality they only have finite spatial volume,
but should still have infinitely many states.  In particular, since
the Ernst instantons of \refs{\Erns,\Gibb,\GaSt} produce \RN\ black holes
slightly above
extremality, they have finite volume, but this
should not necessarily mean that there can be only
finitely many states produced.  The infinite number of states are described
by including fluctuations about the instanton, and arise from the infinite
volume in $\rst$.  (Ref.~\BOS\ argues that finite volume means that there
are only finitely many states.)
However, as argued above, the
fluctuations describing the production of the infinite states destroy
the instanton, and so a definite conclusion cannot be drawn; production
could well be supressed.  The role of
the massless excitations and their backreaction is thus essential.  The
finiteness of the volume alone does not directly imply a finite rate.
It should be noted that \BOS\ also suggested the idea that any attempt
to accelerate dilatonic black holes would excite them, and
advocated the view that
effective field theories are therefore not useful in describing them.
However, it is not clear
that this happens in the dilatonic case.  If one ignores the s-wave
fermions, the excitation spectrum about the throat has a
gap\refs{\HoWi,\DXBH}.  Therefore in the absence of fermions it is
plausible that one could accelerate
one of these objects without excitation.  However, a detailed study of
this problem is difficult due to infinite growth of the coupling in the
vicinity of the black hole; as a whole the proposal also
founders on the rocky
shoals of strong coupling.

\newsec{Discussion}

Remnants and certain types of information loss have been argued to be
different views of the same scenario; if the remnants are truly invisible
then information is lost.  This suggests versions of information loss that
do not violently heat the low-energy world.  However, these
suffer the same difficulties as real remnants, namely the problem of
infinite loss of energy to the Hilbert space of remnants.

\RN\ black holes suggest a possible paradigm for a successful remnant
theory.  Assuming that information is not re-emitted in Hawking radiation,
they should have infinitely many states yet they are not observed to be
infinitely produced.  This paper has made an attempt to understand how this
can happen.  In particular, it is suggested that the couplings of these to
external fields are very non-minimal, and could depend sensitively on the
internal state of the remnant.  Such dependence is essential to eliminate
infinite production.  Unfortunately a detailed model of such dependence has
not yet been found, although it is strongly suggested by the instanton
calculations.  Therefore this paper only represents a suggestion of
how the infinite production problem might be solved.

Since \RN\ black holes do appear to
offer an example of a theory in which objects have
an infinite number of internal states, yet are not infinitely produced, then
it is easy to imagine abstracting the essential features to a theory of
Planck-scale remnants for neutral black holes.  The possibility of there
existing such a theory solves the black hole information paradox.  It
would of course still be extremely interesting to explore how one could get
such a theory of remnants out of quantum gravity.

Other issues that should be confronted if such a theory is to solve the
information conundrum are those of CPT and black hole thermodynamics.  In
particular, in the former context the \RN\ paradigm seems to suggest that
there should be both ``white'' and ``black'' remnants which are CP
conjugates.  One would also like to
understand the connection between remnants and the
second law of black hole thermodynamics.  If information is not returned in
Hawking radiation it is difficult to interpret the Bekenstein-Hawking
entropy as corresponding to the number of states inside a black hole.
Another possibility is that the entropy
indicates the amount of information that can be lost to the internal
remnant state during the course of formation and evaporation of a black
hole from an initial mass $M$.  In this context the real entropy of the
black hole is much larger than given by the Bekenstein-Hawking formula, as
the hole could have been formed by evaporation from a much larger hole.
Furthermore, apparent violations of the second law could be imagined from
dropping such small black holes into big ones.  Perhaps the second law is
only valid in a limited domain, and the Bekenstein-Hawking entropy places
bounds on information transfer rather than information content.

\bigskip\bigskip\centerline{{\bf Acknowledgements}}\nobreak
This work was supported in part by DOE grant DOE-91ER40618 and
by NSF PYI grant PHY-9157463.  I wish to thank T.~Banks, K.~Kucha\v r, S.
Hawking,
G.~Horowitz, D. Lowe,  J.~Polchinski, A.~Strominger,
L.~Susskind, and S. Trivedi for helpful conversations.  I would
particularly like to thank M.~Srednicki for valuable suggetions.

\listrefs

\end